\documentclass[%
 reprint,
 amsmath,amssymb,
 aps,
]{revtex4-1}

\usepackage{graphicx}% Include figure files
\usepackage{epstopdf}% Makes eps files work
\usepackage{dcolumn}% Align table columns on decimal point
\usepackage{bm}% bold math
\begin{document}

\preprint{APS/123-QED}

\title{Towards an experimental demonstration of some novel qubit behavior}
%\title{Quantum phase statistical inference algorithms as a means of going beyond the Heisenberg limit in a standard interferometer}
%\title{Quantum Phase Representation of SU(2) Interferometry}
%\title{Vacuum Entangled N00N States for Tracking Within a Minimally Resourced Quantum Phase Estimator}
% Force line breaks with \\
%

\author{Scott Roger Shepard}
\affiliation{
Department of Electrical Engineering and Computer Science, College of Engineering \&
Science\\
Louisiana Tech University, Ruston, LA 71272, USA\\
}


\begin{abstract}
We show that the quantum angle measurement  for x-polarized photon number states results in an angle which will {\it never} correspond to the y-axis for an odd number of photons; yet for an even number of photons it {\it always} can. The analogy of this surprising effect for other particles or ions (qubits in their spin up state) is presented and we show why such an effect cannot be observed in a standard SU(2) interferometer.  The quantum phase representation of such an interferometer provides clues as to how its apparatus might be modified in order to demonstrate these phenomena. The same representation is then used 
 to provide insight on the range and 
%\mbox{sensitivity} 
sensitivity that one can expect from quantum phase 
statistical inference 
algorithms
% which utilizes only coherent state interferometery and yet can surpass the ``Heisenberg limit'' by  orders of magnitude.
% !
 which  can 
%substantially 
surpass the ``Heisenberg limit.''
% by  orders of magnitude.

\begin{description}
\item[PACS numbers]
42.50.St, 42.50.Dv, 42.50.Ex, 42.50.Lc
\end{description}
\end{abstract}

%Use showkeys class option if keyword
\maketitle

%\tableofcontents

%%%%%%%%%%%%%%%%%%%%%%%%%%%%%%%%%%%%%%%%%%

\section{Introduction}

The general theory of the quantum measurement of relative-phase is summarized in [1] wherein it is also shown to be equivalent to the quantum measurement of an angle about an axis (complementary to the angular momentum along that axis) when the relative-phase is taken to be between the ``up''and ``down'' oscillators of Schwinger's harmonic oscillator model of angular momentum [2]. The quanta of Schwinger's oscillators are presumably unphysical (as they are spin-1/2 primitives which behave like bosons) yet that model has proved  useful for a variety of visualizations and calculations --- including the analysis of beam splitters and interferometers in quantum optics [3]. A slight modification of  Schwinger's harmonic oscillator model of angular momentum permits the inclusion of photons [4] wherein the primitives become the physically realizable quanta of the right and left circular polarization modes of an electromagnetic plane wave. Therein the quantum angle measurement (equivalent to the relative-phase measurement between the two circular polarization modes) provides novel insight on optical polarization; including the fact that for x-polarized photon number states such angle will {\it never} correspond to the y-axis for an odd number of photons, but for an even number of photons it {\it always} can [1], as demonstrated in Fig. 1 for the cases of one, two and three x-polarized photon number states.  
Owing to a factor of two between Schwinger's model and its photonic variation [1,4] the analogy of this surprizing effect, for fermions and ordinary (i.e., non-photonic) bosons, is that ``spin up can point down'' for bosons but it never does for fermions (as demonstrated in Fig. 2 for spin-1/2, 
spin-1
%\mbox{ spin-1} 
and spin-3/2 particles);
%\hspace{.1in} 
wherein by ``spin up along $\bar{x}$''   we mean $m_{x}=j$ where $m_{x}$ is the eigenvalue of $\hat{J}_{x}/ \hbar$ for a particle of total angular momentum labeled by $j$; and by ``point down'' we mean the angle about the z-axis, $\phi$, is $\pi$ (where $\phi=0$ corresponds to the x-axis). Note in passing that Fig. 2 also demonstrates the phenomenon of ``spin up really does point up'' [4] i.e., not only can $\phi=0$ occur for such states, but that is also its most likely value --- in contrast to a vector model which (prior to the establishment of the quantum angle formalism) would attempt to {\it  infer} $\phi$  from the measurement of $\hat{J}_{x}$ and $\hat{J}^{2}$ (rather than directly describing the {\it measurement} of $\phi$). Returning to the ``spin up can point down'' for bosons phenomenon: note that the angular distributions in Fig. 2 are also the relative-phase distributions between the two oscillators of Schwinger's model. Since such oscillators can also model the upward and downward paths in an interferometer the following question naturally arises. Namely, can the surprizing behavior at $\phi = \pi$ in Fig. 2 (analogous to the  surprizing behavior at $\phi = \pi/2$ in Fig. 1) be observed in a standard SU(2) interferometer? 
The answer turns out to be no. 

The quantum phase representation of an SU(2) interferometer discussed herein shows that there is no inconistancy and illucidates the relationship between these two quantum measurements. 
%(that of quantum phase and that of an interferometer).
It  will also fascilitate  the calculation of some of our results   
as well as provide 
%{a visualization of how an interferometer works.}
%novel visulaizations and
 insight on the range and 
%\mbox{sensitivity} 
sensitivity that one can expect from a quantum phase \mbox{estimation}
 %\mbox{algorithm} 
algorithm
% which utilizes only coherent state interferometery and yet can surpass the ``Heisenberg limit'' by  orders of magnitude.
% !
 which  can surpass the ``Heisenberg limit'' by  orders of magnitude while 
% merely 
utilizing
%coherent states within a 
%standard 
the simple apparatus of an
SU(2) interferometer!
%[redo: qm knowledge...resources...to complement words in next paragraph]

%%%%%%%%%%%%%%%%%%%%%%%%%%%%%%%%%%%%%%%%%%%%%%

%\begin{figure}[h]
\begin{figure}[t]
\centering 
\includegraphics[scale=.85]{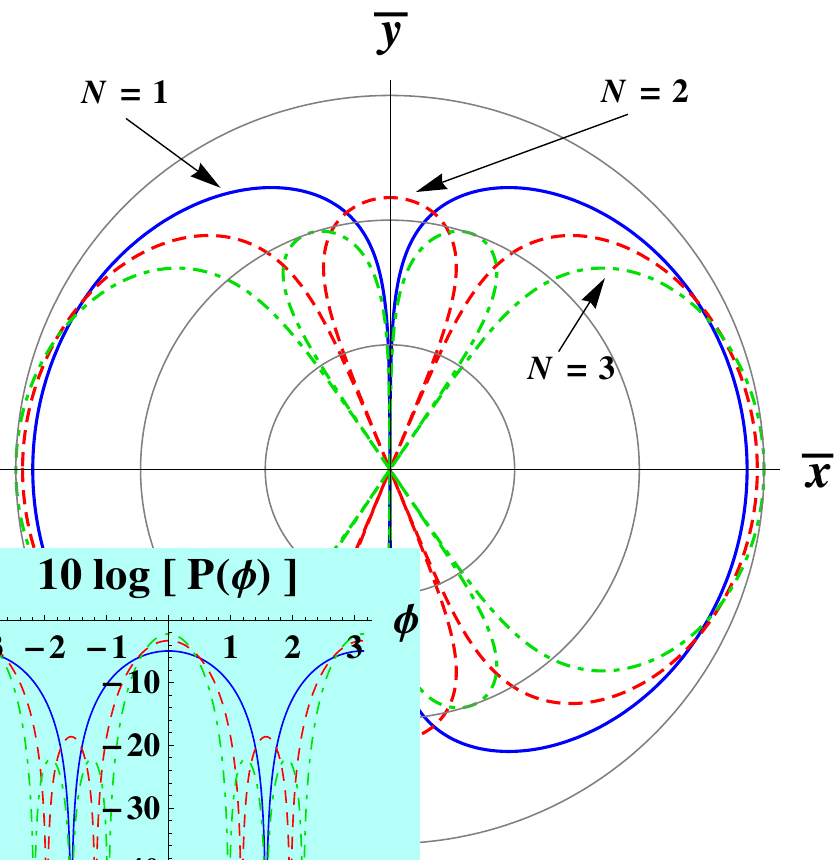}
\caption{(color online).
Quantum angle distribution about the z-axis for x-polarized photon 
number states of: N = 1 (solid, blue); N = 2 (dashed, red); and N = 3 
(dot-dashed, green). Concentric gray circles deliniate steps of 20 
dB. The inset shows the same information on a semi-log plot.}
\end{figure}

%%%%%%%%%%%%%%%%%%%%%%%%%%%%%%%%%%%%%%%%%%%%%%%%

%THIS FILLER PARAGRAPH IS JUST A NOTE TO MYSELF In section II we ... qphirep...and POM the adding amplitudes
%In section III we ... turn out attention/focus...single port number states ...compare to examples of other ways of calculating (or %just "examples") ...and show that Poisson wieghted = coherent state (single port)...and noon...sensitivity, feasibilty  and %range...algorithm...section...results orders of magnitude beyond HL, essentially error-free...resiliance to AWGN

%%%%%%%%%%%%%%%%%%%%%%%%%%%%%%%%%%%%%%%%%%%%%%%%%%%%%%

\section{quantum angle distributions for  linear polarized photon number states}

The transformation between linear and circular polarization basis annihilation operators follows that of their associated classical vectors:
 \begin{eqnarray}
 \sqrt{2} \  \hat{a}_{r} & \equiv  \hat{a}_{x} +  i  \hat{a}_{y} \hspace{2mm} \text{and} \hspace{2mm} \sqrt{2} \ \hat{a}_{l} &  \equiv  \hat{a}_{x} -  i  \hat{a}_{y},  \ \text{so that} 
\nonumber \\
 \sqrt{2} \ {\hat{a}_{x}} & = \hat{a}_{r} + \hat{a}_{l} \hspace{2mm} \text{and}  \hspace{2mm} i \sqrt{2} \ {\hat{a}_{y}} &  = \hat{a}_{r} - \hat{a}_{l}.
 \end{eqnarray}
Thus one x-polalrized photon $|1 \rangle_{x} |0 \rangle_{y}$ is a supperposition of one right-handed ($r$) circularly polarized photon and one left-handed ($l$) circularly polarized photon, which have been created ``in phase'' (which we define to mean that there is no phase shift between the creation operators which create the two modes) i.e.,
 \begin{eqnarray}
 & \hat{a}_{x}^{\dagger} |0,0 \rangle &  = |1 \rangle_{x} |0 \rangle_{y} = (\hat{a}_{r}^{\dagger}+\hat{a}_{l}^{\dagger}) |0,0 \rangle/ \sqrt{2}
\nonumber \\
& & =  (|1 \rangle_{r} |0 \rangle_{l} + |0 \rangle_{r} |1 \rangle_{l})/ \sqrt{2}.
\end{eqnarray}
 Throughout this paper we  adhere to the above definiiton of two modes being ``in phase,'' which is not to be confused with the probability of the outcome of zero when their relative-phase is measured.  

%Throughout this paper we also define to meaning of the words ``out of phase'' to be that two modes are out of phase when there %is phase shift of $\pi$ between the creation operators which create those two modes. This is not to be confused with the %probability of the outcome of $\pi$ when their relative-phase is measured --- which we  refer to instead as ``out of phase type %behavior.''

%\vspace{1in}

To form the quantum angle representation of such a state we use the photonic primitives [1,4] (instead of Schwinger's [2] primitives) which reflect the fact that a photon is a ``spin-1 particle with $m = 0$ missing'' [5] wherein
\begin{equation}
j \rightarrow j_{p} \equiv n_{r} + n_{l} \hspace{2mm} \text{and}  \hspace{2mm} m \rightarrow m_{p} \equiv n_{r} - n_{l}.
\end{equation}
In general, as summarized in [1], we can perform the quantum angle measurement via  a ``snapshot'' in absolute time (a conditional probability operator measure, which involves the addition of amplitudes associated with different $j$) or via a ``time average''
% over absolute time 
(a marginal probability operator measure, which involves the addition of probabilities associated with different $j$). For states with a unique value of $j$ however (such as the x-polarized photon number states) both of these procedures reduce to a simple Fourier transform of the complementary wavefunction $\psi_{j,m} \equiv \langle j,m| \psi \rangle$, the magnitude square of which yeilds the quantum angle distribution, viz.,
\begin{equation}
\psi^{(j)}(\phi) = \sum_{m=-j}^{j} \psi_{j,m} e^{  i m\phi} 
\hspace{1mm} \text{and}  \hspace{1mm}
 P^{(j)}(\phi) = \frac{ |\psi^{j}(\phi)|^2}{2 \pi}.
\end{equation}
%
%%%%%%%%%%%%%%%%%%%%
%some day footnote that minus into phi (was history) but want phi rep to rotate right 
%but NOT fixing that here!! oh, go ahead!!
%%%%%%%%%%%%%%%%%%%%%%%%%%%%%%%%%
%
The quantum angle distribution of a single x-polarized photon utilizes (3) so that $m \rightarrow m_{p}$ increments by two and we obtain 
\begin{equation}
P^{(j_{p}=1)}(\phi) = \frac{1}{2 \pi} [\sqrt{2} \ \text{Cos}(\phi)]^2.
\end{equation}
%The superscript will now be taken to be $j$=N...no, when interf. we'll want it to be j again when do coh.st. 
Similarly, we have
%\begin{eqnarray}
 %|2 \rangle_{x} |0 \rangle_{y}  =  \hat{a}_{x}^{\dagger}|1 \rangle_{x} |0 \rangle_{y}/ \sqrt{2}  = %(\hat{a}_{r}^{\dagger}+\hat{a}_{l}^{\dagger}) |1 \rangle_{x} |0 \rangle_{y}/2  & &
%\nonumber \\
%\hspace{-1 in}= \frac{1}{\sqrt{2}} (\frac{1}{\sqrt{2}} )^2 (\sqrt{2} |j_{p}=2, m_{p}=2 \rangle  & &
%\nonumber \\
%+ 2 |j_{p}=2, m_{p}=0 \rangle + \sqrt{2} |j_{p}=2, m_{p}=2 \rangle) & &
%\end{eqnarray}
\begin{eqnarray}
%\begin{equation}
 |2 \rangle_{x} |0 \rangle_{y}  =  \hat{a}_{x}^{\dagger}|1 \rangle_{x} |0 \rangle_{y}/ \sqrt{2}  = (\hat{a}_{r}^{\dagger}+\hat{a}_{l}^{\dagger}) |1 \rangle_{x} |0 \rangle_{y}/2  & &
\nonumber \\
\hspace{-1 in}= \frac{1}{\sqrt{2}}  \left( \frac{1}{\sqrt{2}}\right)^2  [ \sqrt{2} \ |j_{p}=2, m_{p}=2 \rangle \ +  & &
\nonumber \\
2 \ |j_{p}=2, m_{p}=0 \rangle + \sqrt{2} \ |j_{p}=2, m_{p}=2 \rangle ] & &
%\end{equation}
\end{eqnarray}
so that the quantum angle distribution of a two-photon x-polarized number state is 
\begin{equation}
P^{(j_{p}=2)}(\phi) = \frac{1}{2 \pi} [1/ \sqrt{2} \ + \ \text{Cos}(2 \phi)]^2.
\end{equation}
In general the x-polarized number states yield
\begin{equation}
P^{(j_{p})}(\phi) = \frac{1}{2 \pi} [ \frac{1}{\sqrt{j!}} (\frac{1}{\sqrt{2} })^{j} \ \{C^{j}_{0} \ + \sum_{m} \  C^{j}_{m} \ 2 \text{Cos}(m \phi)\}]^2
\end{equation}
where $j \rightarrow j_{p}$ is always an integer and $m \rightarrow m_{p}$ so that the sum is from 1 when $j_{p}$ is odd (in which case 
$C^{j}_{0} \equiv 0$)  or from 2 when $j_{p}$
 is even. The sum is to an upper limit of $ j_{p}$ in increments of two; and the coefficients are generated recursively from
\begin{equation}
C^{j}_{m} = \sqrt{\frac{j+m}{2}}\  C^{j-1}_{m-1} + \sqrt{\frac{j - m}{2}} \  C^{j-1}_{m + 1}
\end{equation}
for $j >1$ with an initial condition of $C^{1}_{1} = 1$ and  $C^{j}_{0} = \sqrt{2 j} \ C^{j-1}_{1}$ for even $j$ (otherwise it vanishes). We note that $C^{j}_{m=j} = \sqrt{j!}$ and 
%at the end of this section
%in the 
%next section
%following 
we will 
later show how these coefficients are related to the Wigner D matricies. 

Note the generalization of ``spin-1 with $m = 0$ missing'' is that $m_{p}$ is odd (even) when $j_{p}$ is odd (even). Therefore (8) reveals that 
 x-polarized photon number states of an odd number have $P^{(j)} (\phi = \pi/2) = 0$ so that the angle will {\it never} correspond to the y-axis for such states. Yet (8) also shows that  x-polarized photon number states of an even number have $P^{(j)} (\phi = \pi/2) \neq 0$ so that the angle {\it always} can correspond to the y-axis for such states.

%do we have to show that phi->2phi? or phi/2 w/ schwinger's? before we say...

%...bosons...

%... rel.phi between 2 modes can be pi when it's meas.d tho these 2 modes are "in phase" ?

A subtle point which now bears mentioning is that the optical polarization basis transformation in (1) belongs to the U(2) group whereas rotations are isomorphic to SU(2) (where the S stands for special in that its unitary matricies also have a determinant equal to one). The beam splitters of section II also belong to SU(2) but notice we restricted our attention to only x-polarized photons in the above (leaving the y-polarization in the vacuum state). Thus we will be able to map to analogous behavior by also restricting our attention to only single-port input excitations of the SU(2) interferometer (leaving the other input port in the vacuum state).  

Formally, the quantum angle measurement for a state with a single value of $j$, e.g., that of a spin-$j$ particle, will require an additional system such as an electromagnetic field to provide a Hilbert space large enough to permit wavefunction collapse (or equivalently a desription of the measurement in terms of sets of commuting observables) [1]. If however the field is in (for example) the vacuum state prior to the measurement then the single-$j$ \mbox{statistics}, so simply presented in (4), will manefest. 
%things such as credit or having a job etc. don't matter 
% it doesn't come from the world of mankind outside of me, it comes from the seed that God has planted inside of me :)

\section{quantum angle distributions for fermions and non-photonic bosons in the spin up state}

To maintain an analogy with the x-polarized photon number states we now consider a spin-$j$ particle which is 
``spin up along the x-axis,'' i.e., a particle in the state $|j,m_{x} = j \rangle$ where $m_{x}$ is the eigenvalue of $\hat{J}_{x}/ \hbar$. Of course we need not employ any oscillators, but in Schwinger's model these states would correspond to up/down oscillators for $\hat{J}_{x}$ (instead of $\hat{J}_{z}$) of photon number $n_{u} = j$ and $n_{d} = 0$ 
(analogous to a single-port input excitation).
%in general 
%single j qta 
%but to keep with the analogy 
%an explicit example will help to mitigate any notational confusion so let's consider 
%j=1, but not yet
We transform the $J_{x}$ representation into the $J_{z}$ representation via a $\pi/2$ rotation about the y-axis
\begin{eqnarray}
|m_{x}\rangle_{x} & = & \widehat{D}_{y} (\pi/2) |m_{x}\rangle = 
\sum_{m=-j}^{j} \langle m|\widehat{D}_{y} (\pi/2)  |m_{x}\rangle  |m\rangle
\nonumber \\
 &= & \sum_{m=-j}^{j} d^{(j)}_{m,m_{x}} (\pi/2) \  |m\rangle,
\end{eqnarray}
where we inserted a resolution of the identity operator, $\hat{I}= \sum_{m=-j}^{j} |m \rangle \langle m|$, on the left of the rotation operator, $\widehat{D}_{y} (\pi/2).$ For kets without a subscript, the z-axis is implied (as it is also implied for $m$ and $\phi$ throughout this paper). For simplicity we  supressed the  $j$ label  in the kets (and bras) since our attention is currently on the single-$j$ statistics
% since these are single-$j$
and the $d^{(j)}_{m,m_{x}} (\pi/2)$ are from the well-known Wigner D matrix elements: 
\begin{equation}
d^{(j)}_{m,m'} (\beta) \equiv \langle m|\widehat{D}_{y} (\beta)  |m' \rangle
\end{equation} 
which can be calculated in a variety of ways [7]. 
%which can be conveniently obtained via 
%\begin{eqnarray}
 %d^{(j)}_{m,m_{x}} (\beta)  & = & \sum_{k} (-1)^{k-m_{x} + m}
%\nonumber \\
 %\frac{\sqrt{(j + m_{x})! (j - m_{x})! (j + m)! (j - m)!}}{(j + m_{x}-k)! k! (j + m - k)! (k - m_{x} + m)!} & &
%\end{eqnarray}
% where $\beta= \pi/2$ in our case
%"maybe dot after [7]. like to keep it simple."
%
%\[ S= \left[ \begin{array}{ccc}
%1 & 2 & 3 
%\\
%1 & 2 & 3 \\
%1 & 2 & 3 
%\end{array} \right]  \hfill (11) \]
%%%%%%%%%%%%%%%%%%%%%%%%%%
%\{beginflushright} (11) \{endflushright}\]
% how absurd \hspace{} will do it but have to do both sides and tweak
%could build an array around my array...
%%%%%%%%%%%%%%%%%%%%%%%%%%%%
%\begin{eqnarray}
%\[ S= \left[ \begin{array}{ccc}
%\left[ 1 & 2 & 3 
%\\
%1 & 2 & 3 \\
%1 & 2 & 3 \right]
%\end{array} \right] \]] 
%\end{eqnarray}
%%%%%%%%%%%%%%%%%%%%%%%%%%
The ``spin up along x'' kets have $m_{x} = j$
% in (10) 
so the coefficients in their $J_{z}$ representation correspond to the first column of the Wigner D matricies
(of $\beta = + \ \pi/2$).

An explicit example will  mitigate the risk of any notational confusion so let's consider spin-one (i.e., $j = 1$) in the following. We have 
\begin{equation}
|1 \rangle_{x} = \frac{1}{2} \ |1 \rangle + \frac{1}{\sqrt{2}} \ |0 \rangle  + \frac{1}{2} \ |-1 \rangle 
\end{equation}
so that (4) yields
\begin{equation}
P^{(j=1)}(\phi) = \frac{1}{2 \pi} [1/ \sqrt{2} \ + \ \text{Cos}( \phi)]^2.
\end{equation}
Notice that the quantum angle distribution of spin-1 up along $\bar{x}$ is related to that of the two-photon x-polarized number state as follows: 
\begin{equation}
P^{(j=1)}(\phi) = P^{(j_{p}=2)}(\phi \rightarrow \phi /2)
\end{equation}
which exemplifies how the surprizing behavior at $\phi = \pi/2$ illustrated in Fig. 1 turns into the surprizing behavior at $\phi = \pi$ illustrated in Fig. 2. Of course, the Wigner D matricies can be calculated via Schwinger's harmonic oscillator model [7] and we could have performed an analysis similar to that which led to (8) via Schwinger's primitives instead of right and left circular polarized photons. Moreover, Schwinger's up/down oscillators can also model the upward/downward paths inside an SU(2) interferometer[8]  and the angle distribution in (4) is also the relative-phase distribution [1] between these two modes. 

\section{relative phase distributions for single-port  photon number state 
%zero degree 
beam splitters }

In the quantum theory of an interferometer [3] there are many possible choices of beam splitter reference planes --- which correspond to many  possible choices of axies of rotation. We choose the y-axis so that the Wigner D matrix elements will occur in what follows. Thus, we can choose  to have either the $J_{x}$ representation ``outside'' the beam splitters and the $J_{z}$ representation ``inside the interferometer'' or vice-versa. We choose the former in order to maintain similarity to the above, as well as to simplify the notation so that we won't need a subscript on $\phi$ since the z-axis will be implied. We also choose the first beam splitter to correspond to a rotation of $+\pi/2$ and the second one to be of $-\pi/2$ so that spin up will always correspond to the upward path.

The transformation of the first beam splitter's 
input annihilation operators (denoted with an ``a'') into its output annihilation operators (denoted with a ``b'') follows that of $d^{(1/2)}_{m,m^{'}}(+ \pi/2)$:
 \begin{eqnarray}
 \sqrt{2} \  \hat{b}_{u} & =  \hat{a}_{u} -   \hat{a}_{d} \hspace{2mm} \text{and} \hspace{2mm} \sqrt{2} \ \hat{b}_{d} &  =  \hat{a}_{u} +  \hat{a}_{d},  \ \text{so that} 
\nonumber \\
 \sqrt{2} \ {\hat{a}_{u}} & = \hat{b}_{u} + \hat{b}_{d} \hspace{2mm} \text{and}  \hspace{2mm}
 \sqrt{2} \ {\hat{a}_{d}} &  = - \ \hat{b}_{u} + \hat{b}_{d}.
 \end{eqnarray}

Thus $ \sqrt{2} \ {\hat{a}_{u}}  = \hat{b}_{u} + \hat{b}_{d}$ (analogous to $ \sqrt{2} \  \hat{a}_{x}  =  \hat{a}_{r} +  \hat{a}_{l}$) splits the single-port input into two ``in phase'' modes inside the interferometer; and we create the $J_{z}$ representation of the input number states (of $J_{x}$ representation) 
in parallel with how we created the circular polarization representation of the linear polarized number states;  but this time, instead of (3) we have  $ j = (n_{u} + n_{d})/2$ and $ m = ( n_{u} - n_{d})/2$, i.e., Schwinger's primitives,  
so that each quanta is associated with a z-component angular momentum  of 1/2. Notice that in (14)
we changed $j$ (and $\phi$, because $m$ changed)  but the coefficients did not change. Since we could also do that calculation via the Wigner D matricies, 
the $C^{j}_{m}$ are an alternate way of calculating the same numbers as the first column of these matricies and we have
\begin{equation}
\frac{1}{\sqrt{(2j)!}} \ (\frac{1}{\sqrt{2} })^{2j} \ C^{2j}_{2m} = d^{(j)}_{m,j} (\pi/2).
\end{equation}
This also proves that the $d^{(j)}_{m,j} (\pi/2)$ must be symmetric in $m$, so their Fourier transform results in terms 
% as in (8) 
 only of the form $\text{Cos} (m \phi)$ and the interpretation of the surprizing effect at $\phi = \pi$ is that
two modes which have been created ``in phase'' (via an \textit{even} number of creation operations) can have a relative-phase of $\pi$ 
when it is measured; yet those created via an {\it odd} number of creation operations can not. 

Note in passing that
a mode input to the $\textit{other}$ port of this beam splitter would be split into two-modes which are created to be 180 degrees out of phase but herein we do not excite that input port, just as we did not bring the y-polarization out of the vacuum state in section II.
So we also need not concern ourselves with the difference between U(2) and SU(2), e.g., the absence of an $i$ in 
$ \sqrt{2} \ {\hat{a}_{d}}   = - \ \hat{b}_{u} + \hat{b}_{d}$
 as opposed to
$i \sqrt{2} \  {\hat{a}_{y}}   = \hat{a}_{r} - \hat{a}_{l}$ since we never used either of these equations; and restricted our attention instead  to 
$ \sqrt{2} \ {\hat{a}_{u}}  = \hat{b}_{u} + \hat{b}_{d}$  and the analogous $ \sqrt{2} \  \hat{a}_{x}  =  \hat{a}_{r} +  \hat{a}_{l}$. Moreover, if we excited some y-polarization (or an input to the 180 degree splitter port) then we would not be surprized to observe an optical polarization angle at $\pi/2$ (or a relative-phase of $\pi$). In the next section we derive the quantum phase representation of an SU(2) interferometer without any restriction on its input state. After which, we will then return to the analysis of the single-port input statistics.

Although they 
%$\textit{do}$ 
correspond to quantum measurements which are realizable (in principle)  no one, to date, has identified an apparatus which can realize the quantum phase or quantum angle measurements. Interferometers are typically used to $\textit{infer}$ phase, so the question arises as to wheather or not such ``out of phase type behavior'' can be demonstrated in an SU(2) interferometer? The answer clearly is no, since if we phase shift the upper path by some amount $\Phi$ and recombine the two via a beam splitter of reference planes associated with a rotation about $\bar{y}$ by $-\pi/2$ then at $\Phi = 0$ the interferometer reduces to the indentity operator --- reproducing the input state at its output $\textit{for all}$ input states $|\psi \rangle$. Thus, spin up at the input can never result in any photons going into the ``down port'' photodetector at the output when $\Phi = 0$ --- independent of the value of $P(\phi = \pi)$.  
%Likewise, at $\Phi = \pi$ the inerferometer simply flips the input state upside down $ \forall |\psi\rangle$.
% --- independent
 
This is not to say that such an effect cannot be observed via some $\textit{other}$ quantum measurement for which we presently  have a physical realization (e.g., perhaps it could be observed
in heterodyne or homodyne detection, etc.).
It is instructive to explicitly demonstrate the relationship between the quantum measurement of phase and the quantum measurement realized in an SU(2) interferometer: 
not only
to demonstrate that there is no inconsistancy in their predictions and provide clues as to how one might modify the interferometer's apparatus
in order 
to observe the surprizing effects above; but also the quantum phase representation of an SU(2) interferometer will provide novel insight as to how an interferometer works (as did the analogies to rotations [3]). 
Most importantly, 
it will 
facilitate proofs
regarding the range and sensitivity one can anticipate in quantum phase statistical inference (QPSI) algorithms --- which can yeild accuracies that are orders of magnitude better than the ``Heisenberg limit,'' while utilizing the simple apparatus of an SU(2) interferometer (and classical signal processing that has been armed with quantum knowledge of the possible outcomes) [8,9].

%%%%%%%%%%%%%%%%%%%%%%%%%%%%%%%%%%%%%%%%%

\begin{figure}
%\centering 
\includegraphics[scale=.85]{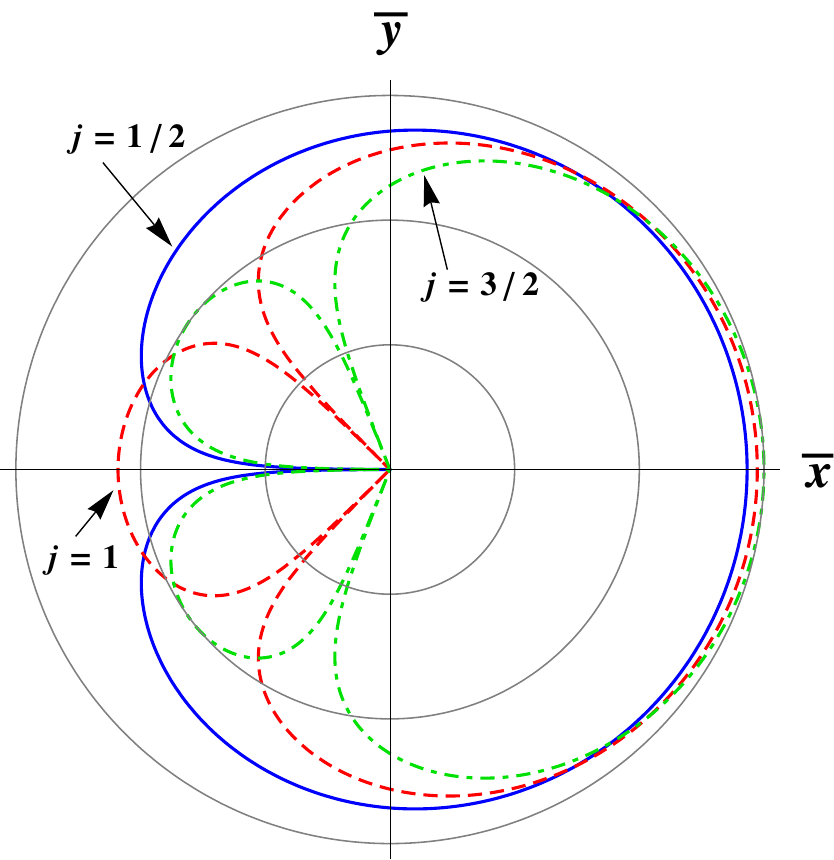}
\caption{(color online). Quantum angle distribution about the z-axis for spin up along $\bar{x}$ \
of: m = j = 1/2 (solid, blue); m = j = 1 (dashed, red); and m = j = 3/2 \
(dot-dashed, green). Concentric gray circles deliniate steps of 20 dB.}
\end{figure}

%%%%%%%%%%%%%%%%%%%%%%%%%%%%%%%%%%%%%%%%%

\section{quantum phase representation of an SU(2) interferometer}

%cut and paste from existing

%%%%%%%%%%%%%%%%%
%do the don't do it twice thing later
%%%%%%%%%%%%%%%

%\vspace{1in}

First, there is a subtle point that we wish to clarify so that the notation is transparent to an audience wider than just the quantum optics community. 
Recall we chose  to have  the $J_{x}$ representation ``outside'' the beam splitters and the $J_{z}$ representation ``inside the interferometer.''
But it would be wrong to say, for example, that with an input state of $|m_{x}\rangle_{x}$ the state of the first beam splitter's output is $\widehat{BS}_{in} \ |m_{x} \rangle_{x}$ and also utilize (10), since that would be tantamount to performing the rotation assosicated with the input beam splitter, $\text{BS}_{in}$, twice. The beam splitter of course $\textit{does}$ perform an operation on these states (in accordance with the transformation of annihitaion operators in the Heisenberg picture, as in (15) for our choice).  The use of (10) makes it look as if we chose to use two different representations of the $\textit{same}$ state --- with its $J_{x}$ representation on the left of $\text{BS}_{in}$ and its $J_{x}$ representation on the right of $\text{BS}_{in}$. 
The subtlety is that those are different \textit{spatial} modes (as clarified in (15) by the use of $\hat{a}$ versus $\hat{b}$). In other words, the  upward/downward paths on the left are different spatial modes than the  upward/downward paths on the right.

%%%%%%%%%%%%%%%%%%%%%%%%%%%%%%%%%%%%%%%%%

\begin{figure}
%\centering 
\includegraphics[scale=.65]{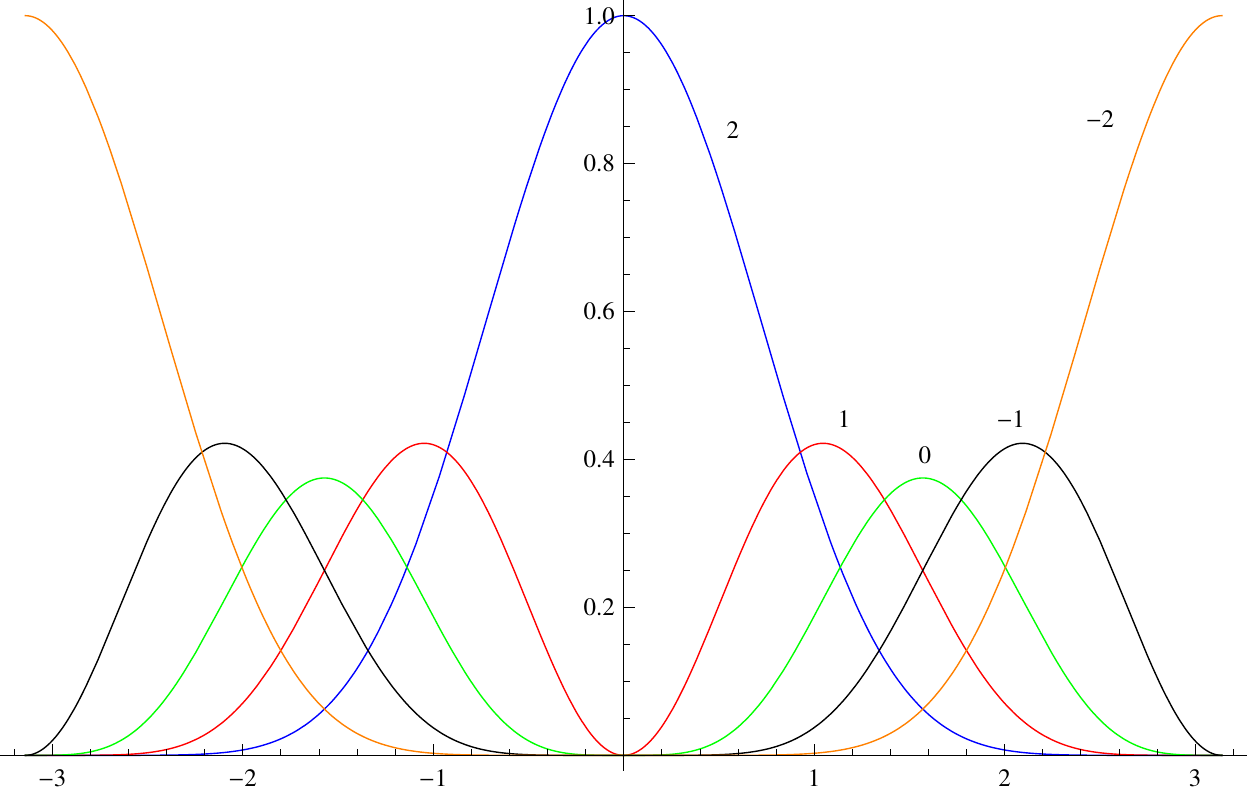}
\caption{(color online). Interferometer measurement statstics $P_{m} (\Phi)$ \ for a single-port four-photon number state input.
Value of  $m$ are indicated from $j$ to $- j$ in the color sequence: Blue, Red, Green, Black, Orange.}
\end{figure}

%%%%%%%%%%%%%%%%%%%%%%%%%%%%%%%%%%%%%%%%%

%%%%%%%%%%%%%%%%%%%%%%%%%%%%%%%%%%%%%%%%%

\begin{figure}[b]
%\centering 
\includegraphics[scale=.65]{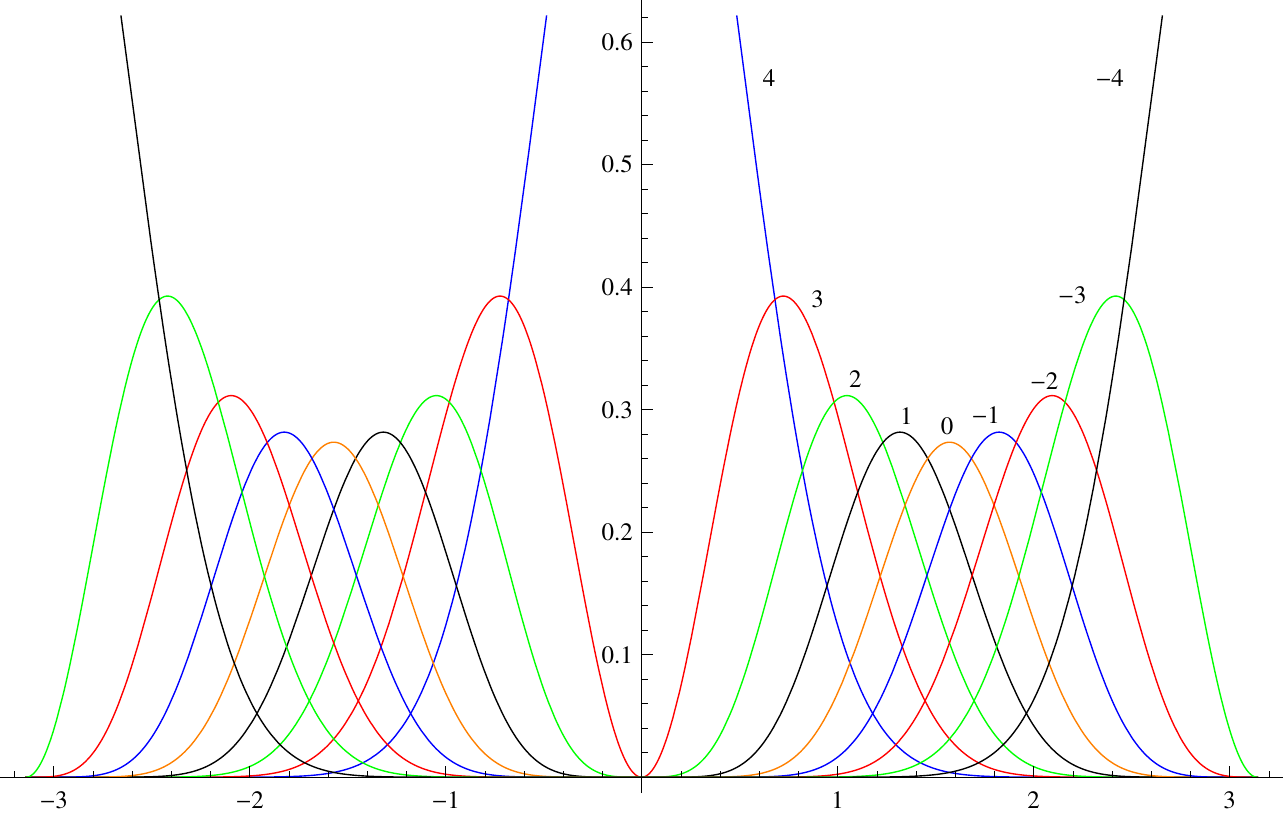}
\caption{(color online). Interferometer measurement statstics $P_{m} (\Phi)$ \ for a single-port eight-photon number state input.
Value of  $m$ are indicated from $j$ to $- j$ in the color sequence: Blue, Red, Green, Black, Orange.}
\end{figure}

%%%%%%%%%%%%%%%%%%%%%%%%%%%%%%%%%%%%%%%%%

Perhaps the safest notation is to let $|\psi \rangle$ represent the state at the output of the first beam splitter (and let  $\text{BS}_{in}$ be thought of as part of the state preperation step --- examples of which in the Heisenberg and Schrodinger pictures being given in the previous sections). Thus, the interferometer's measurement statistics can be written as 
$\text{P}^{(j)}_{m'} (\Phi) =  |\Psi^{(j)}_{m'}(\textbf{$\Phi$})|^2 $ where

\begin{equation}
\Psi^{(j)}_{m'} (\Phi)\equiv \ _{x}\langle m' |\widehat{D}_{z}(\Phi) |\psi \rangle
\end{equation}
is the underlying wavefunction in measurement outcome $m' = (n_{u} - n_{d})/2$. Note that $m'$ (the eigenvalue of  $\hat{J}_{x}/ \hbar$) is 1/2 the difference of the number of photodetection events in the up/down photodetectors (within some observation time $T$) when one imposes a differential phase shift $\Phi$ between the upward and downward paths inside the interferometer  --- analogous to the rotation $ \widehat{D}_{z}(\Phi)$. 
For arbitrary states (rather than those comprised of a single value of $j$, such as the single-port photon number states) we could argue that the  measurement outcomes from different $j$ are distinguishable in principle (wheather or not we record $j$) so that in a standard interferometer we would add probabilities: $\text{P}_{m'} (\Phi) = \sum_{j} \text{P}^{(j)}_{m'} (\Phi)$ 
[10].

Notice that we chose a ``$+$'' in the Fourier transform going $\textit{into}$  $\phi$ in (4) so that right-handed rotations displace $\phi$ to the right: 
\begin{equation}
\langle \phi | \widehat{D}_{z} (\delta \phi) |\psi \rangle = \psi(\phi -\delta \phi) 
\hspace{2mm} \text{since} \hspace{2mm} 
\end {equation}
\begin{equation}
\psi(\phi) \equiv  \langle \phi | \psi \rangle
\hspace{2mm} \text{so} \hspace{2mm}
\langle \phi |  =  \sum_{m}  e^{ + i m \phi} \langle m|
\hspace{2mm} \text{and} \hspace{2mm}
\end{equation}
\begin{eqnarray}
\langle \phi | \widehat{D}_{z} (\delta \phi) & \longleftrightarrow & \widehat{D}_{z}^{\dagger} (\delta \phi)  | \phi \rangle
\nonumber \\
& = &
 e^{+ i \hat{J}_{z} \delta \phi / \hbar} \sum_{m}  e^{ - i m \phi } |  m \rangle 
\nonumber \\
& = & |  \phi - \delta \phi \rangle.
\end{eqnarray}

Inserting a resolution of the identity operator in terms of the angle-kets, 
$\hat{I}=\int_{-\pi}^{\pi}  \frac{d\phi}{2\pi} \ | \phi\rangle\langle\phi|$ ,
on the left of the rotation operator in (17) then yields
\begin{eqnarray}
\Psi_{m^{'}}(\Phi) & \equiv &
 \ _{x}\langle m' | \widehat{D}_{z}(\Phi) |\psi \rangle
\nonumber \\
& = & \ _{x}\langle m' |
\int_{-\pi}^{\pi}\frac{d\phi}{2\pi} \ |\phi\rangle \ \psi(\phi-\Phi).
\end{eqnarray}
Now, utilizing the adjoint of (10) we have
\begin{eqnarray}
\ _{x}\langle m' | \phi \rangle & = & \sum_{m} \ _{x}\langle m'| m \rangle \ e^{- i m \phi}
=
\sum_{m}\langle m^{'}| \widehat{D}_{y}^{\dagger}(\pi) | m \rangle \ e^{- i m \phi}
\nonumber \\
& = & \sum_{m} d_{m',m}^{(j)}(- \pi/2) \ e^{- i m \phi} \equiv S^{(j)}_{m'} (- \phi)
\end{eqnarray}
so that the quantum phase representation of an SU(2) interferometer is
\begin{equation}
\Psi^{(j)}_{m}(\Phi)=\int_{-\pi}^{\pi}\frac{d\phi}{2\pi} \ 
%\[ 
S^{(j)}_{m}(-\phi) \ \psi^{(j)} (\phi - \Phi) 
%\]
\end{equation}
where the functions $S^{(j)}_{m}(\phi)$ are the Fourier transforms of the m-th row of the Wigner D matricies (at $\beta = - \pi/2$). 

The $S^{(j)}_{m}(\phi)$ functions represent the action of the output beam splitter and its two photodetectors; as they project the angle-kets onto the kets of what the interferometer measures.  
%(for splitter) 
These state-independent functions are known (in as much as the $d_{m, m'}^{(j)}$ are) and the only dependence of the interferometer's measurement statistics on the quantum state of the field is via their convolution with its angle representation (which reduces the action of the rotation operator to a simple translation of a functional argument).

It is easy to show that the $S^{(j)}_{m}(\phi)$ form a set of orthogonal functions (for any $j$) 
\begin{equation}
\text{let} \ \zeta  \equiv  \int_{-\pi}^{\pi}\frac{d\phi}{2\pi} \ 
S^{(j)}_{m}(\phi) \ S^{(j)}_{n}(\phi). 
%\nonumber \\
\end{equation}
So, from the orthogonality of the $e^{i m \phi}$
\begin{eqnarray}
 \zeta & = & \int_{-\pi}^{\pi}\frac{d\phi}{2\pi} \ 
\left( \sum_{m'} d^{(j)}_{m,m'} \  e^{i m' \phi} \right) \left( \sum_{m''} d^{(j)}_{n,m''} \  e^{i m'' \phi} \right)
\nonumber \\
 & = & \sum_{m'}\left( d^{(j)}_{m,m'} \right)  \left( d^{(j)}_{n,m'} \right)
\end{eqnarray}
where the sum is over the columns of ``row m'' times ``row n'' (and we supressed $\beta = -\pi/2$). But the n-th row of $d^{(j)}_{n,m'}$ is also the n-th column of its inverse, hence $\zeta = \delta_{m,n}$ (the Kronecker delta) as asserted.   
%...equation...put Fig. 3 top left of this page...

When a single-port number state is input to the interferometer (16) shows that $\psi(\phi)$ itself becomes $S^{(j)}_{m=j}(\phi)$ so %that
\begin{eqnarray}
\Psi^{(j)}_{m=j}(\Phi) & = & \int_{-\pi}^{\pi}\frac{d\phi}{2\pi} \ 
%\[ 
\psi(-\phi) \ \psi (\phi - \Phi) 
%\]
\nonumber \\
 & \rightarrow & \int_{-\pi}^{\pi}\frac{d\phi}{2\pi} \ S^{(j)}_{m=j}(-\phi) \ S^{(j)}_{m=- j}(\phi)
%\nonumber \\
%\textit{as} \Phi & \rightarrow & \pi
\end{eqnarray}
as $\Phi \rightarrow \pi$, where we used $m \rightarrow - m$ under a rotation by $\pi$. 
Orthogonality then shows 
%$\Psi^{(j)}_{m=j}(\Phi = \pi) = 0$ 
that this vanishes, independent of the value of $\psi(\phi = \pi)$. 

Thus there is no inconsistency. The state-independent behavior of the interferometer at the extremes of $\Phi = 0$ and $\Phi = \pi$ can of course been seen more readily by considering it as one equivalent rotation, but now we have clues as to how we might modify the apparatus in order to reveal the behavior at $\phi = \pi$.
It is this squaring of the Wigner D matrix elements which prevents the observation of such effects and we make the conjecture that instead of ``beating the signal with itself'' (so to speak) we should beat it with a seperate local oscillator (as in heterodyne detection). An analgous conjecture would be that these effects might be observed in scattering experiments on trapped ions (e.g., perhaps as in  Raman heterodyne spectroscopy). 

We now turn our attention to the use of this  representation  
 to provide insight on the range and 
%\mbox{sensitivity} 
sensitivity that one can expect from quantum phase 
statistical inference 
algorithms [8,9]. 
The squares of the $d^{(j)}_{m=j, m'}$ turn out to be proportional to the binomial coefficients, thus we can prove that for the single-port number states we have $P^{(j)}_{m = j}(\Phi) = \left[ \text{Cos}(\Phi /2)\right]^{4j}$. Similarly we can show that 
\begin{equation}
P^{(j)}_{m = j - 1}(\Phi) =2 j  \left[ \text{Cos}(\Phi /2)\right]^{4 j -2}  \left[ \text{Sin}(\Phi /2)\right]^{2}
\end{equation}
etc.,  lowering in the same fashion down to $P^{(j)}_{m = - j}(\Phi) = \left[ \text{Sin}(\Phi /2)\right]^{4j}$. Closed-form expressions are useful but so are geometric interpretations. 

Fig. 3 presents the $P^{(j)}_{m}(\Phi)$ versus $\Phi$ for $j=2$. The behavior of $P^{(j)}_{m = j}(\Phi)$ stems from (23) as it is the phase representation of the state convolved with itself. Let $m \ge 0$ for the moment.
As $m$ decreases towards zero the higher frequency components in $S^{(j)}_{m}(\phi)$ increase in amplitude (because the  $d^{(j)}_{m, m'=j}$ get larger in absolute value) so that their convolution with the phase representation will displace the resulting peaks further away from $\phi = 0$. The behavior for $m < 0$ follows from a rotation by $\pi.$ When we consider e.g., a simultaneous LMS fitting of these functions to measured interferometer statistics (obtained during a fixed and unknown value of $\Phi$)  
we see that the $\textit{near}$ monotonicity of the $P^{(j)}_{m = \pm j} (\Phi)$ on [0, $\pi$] will resolve which side of the peaks we must be on for the other statistics; which in turn enable decent sensitivity (slope) throughout this range. 
 Fig. 4 illustrates similar behavior for $j = 4$ wherein we
also see the increased sensitivity resulting from the use of more photons; as well as the anticipation of good performance closer to zero and $\pi$ (where all slopes must vanish due to symmetry). 

% thank you Lord :)

%\vspace{.15in}
%\hspace{-0.13in}{\textbf{Acknowledgments}}
%\vspace{.15in}

%\noindent S.R.S. would like to acknowledge support from NASA.

%%%%%%%%%%%%%%%%%%%%%%%%%%%%%%%%%%%%


\begin{thebibliography}{99}
\bibitem{10} As designers of quantum measurements however one could envision a ``stroboscopic interferometer'' in which we would add the amplitudes for outcomes $m'$ of different $j$. Indeed, such a measurement would be complementary to the ``snapshot'' angle measurement [1].
\end{thebibliography}
\end{document}